# Impact of Angle Misalignment on the Performance of a combined Optical and millimeter-wave Transceiver enabled by a pair of Optical Harmonically Locked Lasers

**Zichuan Zhou, Amany Kassem, Zun Htay, Izzat Darwazeh and Zhixin Liu**
*Department of Electronic and Electrical Engineering, University College London (UCL), United Kingdom*
*Author e-mail address: zczlzz0@ucl.ac.uk*

**Abstract:** We demonstrated combined free-space optics (FSO) and D-band(110-170GHz) millimeter(mm-wave) transceiver enabled by precisely locked lasers with low phase noise. Combined capacity and tolerance to angle misalignment are studied using 100Gb/s optical and mm-wave signals. © 2025 The Author(s)

## 1. Introduction

The increasing demand for capacity in satellite communication and wireless backhaul has been driving the development of high-capacity point-to-point wireless transmission [1-3]. Free-space optics (FSO) transceivers, as key enablers, have demonstrated 1 Tbps FSO transmission between observatories with large elevation differences, showing strong potential for GEO-satellite feeder link [3] and wireless connection between radio units [2]. Unfortunately, FSO transmission is highly sensitive to angular misalignment due to its highly directional beam, thus requires complex and costly beam alignment and control system [4]. Meanwhile, millimeter-wave (mm-wave), particularly D-band systems (110-170 GHz), are emerging as promising candidates for high-capacity wireless links. They offer up to 60 GHz unlicensed bandwidth with increasingly mature components/subsystems, such as D-band mixers, amplifiers and antennas [5]. Compared to FSO, mm-wave transceivers exhibit significantly higher tolerance to beam misalignment. However, their performance is constrained by carrier phase noise, which increases quadratically with frequency in conventional electronic frequency synthesizers [6]. In addition to the complementary performance between misalignment tolerance and capacity, FSO and mm-wave also exhibit complementary weather-dependent performance. Specifically, FSO is tolerant to rainfall but is severely attenuated by fog and cloud, whereas mm-wave signals penetrate fog and cloud effectively but experience strong scattering losses in rain [7].

The above has led to investigations of combining both FSO and mm-wave transceivers for high capacity and resilient transmission. In [8], researchers have demonstrated combined FSO and mm-wave transmission at carrier frequency of 80GHz, in what is fundamentally two separate systems requiring three lasers, one for FSO signal generation and two for mm-wave signal generation based on optical heterodyne, significantly increasing the cost and system complexity. Moreover, the two lasers used for mm-wave signal generation are not frequency and phase locked, which fundamentally limits the phase noise performance of transmitted mm-wave signal. In [9], researchers have proposed and demonstrated a combined FSO and mm-wave transmitter with mm-wave signal generated by directly modulating laser with intermediate-frequency (IF) signal, followed by an electronic frequency up-convertor, which converts IF to desired mm-wave frequency. Compared to [8], the required number of lasers is reduced from three to two. However, the mm-wave signal performance is fundamentally limited by the electronic local oscillator phase noise, which scales as frequency increases [6]. Recently, a shared FSO and mm-wave transmitter architecture has been proposed, which is based on the incomplete photoelectric conversion of uni-travelling-carrier photodetector (UTC-PD) [10]. However, as same UTC-PD simultaneously outputs both FSO and mm-wave signal, it is challenging to further amplify FSO and mm-wave signal after UTC-PD, fundamentally limiting the transmission distance. Additionally, the proposed scheme only allows same data to be transmitted in both FSO and mm-wave channel, prohibiting dynamic channel resource allocation under various channel conditions.

In this work, we propose and demonstrate a combined FSO and mm-wave transceiver featuring ultra-low phase noise mm-wave generation and optical modulation using a pair of optically phase locked lasers [11]. Two ultra-low linewidth lasers are frequency and phase locked to generate high-quality D-band mm-wave carrier, which is fed into a fundamental mixer for D-band signal generation. One of the ultra-low linewidth lasers is separately modulated and fed into collimator for FSO transmission. Using the proposed scheme, we analyzed the alignment tolerance and overall system performance achieving 100Gbps 64QAM transmission for both FSO and D-band mm-wave links over 55cm. Compared to FSO transceiver alone, the proposed combined FSO and D-band mm-wave transceiver improves the misalignment tolerance by a factor of 193 and achieves a 2.2 dB coherent combining SNR gain, demonstrating a robust alignment resilience, high data capacity and phase-coherent signal enhancement within a single architecture.

## 2. Experimental Setup

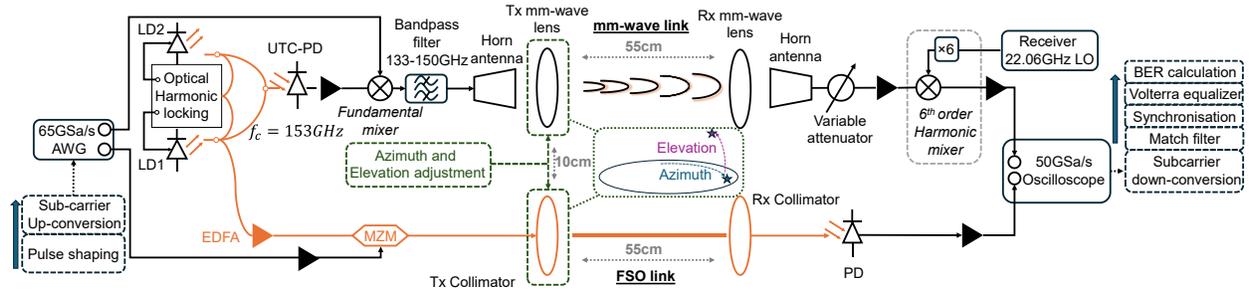

Fig. 1. Experimental setup for combined FSO and mm-wave D-band transmission system

Fig. 1 shows the experimental setup for combined FSO and mm-wave D-band transmission system. The IF signal was generated using a 65GSa/s arbitrary waveform generator (AWG) with 25GHz bandwidth and 8-bit digital resolution. A pseudorandom binary sequence (PRBS) of $2^{20}-1$ length was used to generate 17GBd 16 and 64QAM digital signals shaped by root-raised cosine pulse filter with a roll-off factor of 0.1, resulting in a baseband signal of 17GHz bandwidth. The baseband signal was then digitally up-converted to a carrier centered at 11.5GHz, resulting in a real-valued doubled sideband sub-carrier-modulated (SCM) signal over 3-20GHz.

At transmitter side, two ultra-low linewidth lasers (i.e. LD1: 7Hz, 16.5dBm and LD2: 40Hz, 15dBm) are used. For mm-wave D-band signal generation, the frequency spacing between LD1 and LD2 is adjusted to 153GHz. The frequency and phase of LD1 and LD2 are locked using optical harmonic locking technique [1,6,11] and then combined and fed into a D-band UTC-PD, outputting a 153GHz carrier with around -8dBm power, which is further amplified by a 20dB gain D-band amplifier. The measured rms jitter (integrated from 10Hz to 10MHz) of generated 153GHz carrier is 14fs [6,11], enabling up to 15dB receiver sensitivity improvement compared to conventional electronic oscillators as demonstrated in [6]. The 153GHz carrier is then fed into D-band fundamental mixer for frequency up-conversion. The IF signal output from an AWG was attenuated to around -13dBm to avoid mixer saturation. The upconverted double sideband mm-wave signal was fed into a 133-150GHz D-band bandpass filter, resulting in >20dB suppression of upper sideband. The output of bandpass filter (around -29dBm) was fed into a D-band horn antenna with 20dBi gain and transmitted through an aspheric lens to a receiver front end with an identical lens-antenna pair spaced 55cm away. The maximum received power measured at antenna output is around -32dBm, corresponding to a total of 3dB D-band link loss. The output of receiver antenna is connected to a D-band variable attenuator, followed by a low noise figure (6.5dB) 13dB gain D-band amplifier and a $6^{th}$ order harmonic mixer based frequency down-converter. The resulting baseband signal is detected by a 50GSa/s 8-bit real-time oscilloscope.

For FSO transmission, an Erbium-Doped Fiber Amplifier (EDFA) amplifies the tapped LD1 output to 15dBm, which is then fed into a Mach-Zehnder modulator (MZM) biased at quadrature point, driven by the amplified output of AWG. The MZM output is fed into transmitter side collimator and detected by same model collimator placed 55cm away from transmitter. The maximum optical power received is around 7dBm, corresponding to 2dB FSO link loss, mainly from optical coupling imperfections. The collimator is connected to a 40GHz photodetector, followed by a 25dB gain RF amplifier before entering the 50GSa/s 8-bit real-time oscilloscope. For both FSO and mm-wave link, the receiver DSP involves digital down-conversion, matched filtering, clock and frame synchronization, and a digital equalizer [12] with 131 first order taps and 3 third-order taps. To study misalignment tolerance, the collimator and aspheric lens are mounted on two separate adjustment stages, allowing up to 6.4- and 4-degree azimuth and elevation adjustment.

## 3. Results

Fig. 2a and 2b show the measured BER with different azimuth and elevation misalignment angles for both FSO and D-band 16/64QAM signal. The hollow symbols and solid symbols represent FSO and D-band signal, respectively. The blue and orange curves represent 16 and 64QAM. When azimuth misalignment angle increases from 3 to 6.4 degrees, BER of 16 and 64QAM D-band slightly degrades due to lower received power. With worst case scenario azimuth misalignment, D-band 16QAM and 64QAM BER are still maintained below HD-FEC (BER=3.8e-3 [13]) and SD-FEC threshold (BER=2e-2 [14]). On the other hand, with azimuth misalignment angle is <0.044 degree, 16 and 64QAM FSO BER outperforms D-band signal. One of possible reasons is the limited bandwidth of D-band components. In contrast, for FSO signals, BER significantly degrades as azimuth misalignment angle increases. For FSO 16QAM and 64QAM, the measured BER increases beyond SD-FEC threshold with 0.044- and 0.033-degree azimuth misalignment. In terms of elevation misalignment tolerance, D-band 16 and 64QAM BER are maintained below HD-FEC and SD-FEC threshold even with 4 degrees of elevation misalignment. While FSO 16 and 64QAM BER increases beyond SD-FEC threshold with 0.0464-degree elevation misalignment. As a result, compared to FSO transmitter, the proposed combined FSO and D-band transmitter improves the azimuth and elevation misalignment tolerance by a factor of 193 (0.033 to 6.4 degree) and 86 (0.0464 to 4 degree) for 17GBd 64QAM transmission. It is worth noting that in this proof-of-concept experiment, the elevation misalignment measurement range is up to 4 degrees, limited by tuning range of adjustment stage. Finally, Fig. 2c shows the 64QAM SNR for three different scenarios: mm-wave D-band only, FSO only and coherent combined FSO-Dband. The red markers represent the coherent combined SNR, and the orange markers represent the FSO-only SNR. For both cases, solid markers indicate azimuth misalignment, while hollow markers indicate elevation misalignment. The blue markers correspond to the D-band SNR, which remains unchanged within 0.06-degree azimuth or elevation misalignment. In this particular measurement, the same data is transmitted on both FSO and D-band channels. The delay difference of received signal is calculated offline, and the synchronized samples are fed into a 2 by 1 equalizer (131 first order taps and 3 third-order taps) for coherent combination. When transmitter and receiver are perfectly aligned, the FSO, D-band and coherent combined signal SNR are 19.2, 18.5 and 20.7dB, respectively (corresponding constellation diagrams shown by insets). This indicates that by coherently combining FSO and D-band signals, we gain 1.5 and 2.2dB SNR improvement compared to FSO only and D-band only cases, which can support higher capacity transmission over longer distance [15]. However, when the azimuth/elevation misalignment angle increases beyond 0.042/0.055 degree, the coherent combined SNR is lower than D-band only case, indicating that coherent combine DSP induced SNR improvement is fundamentally limited by FSO misalignment tolerance.

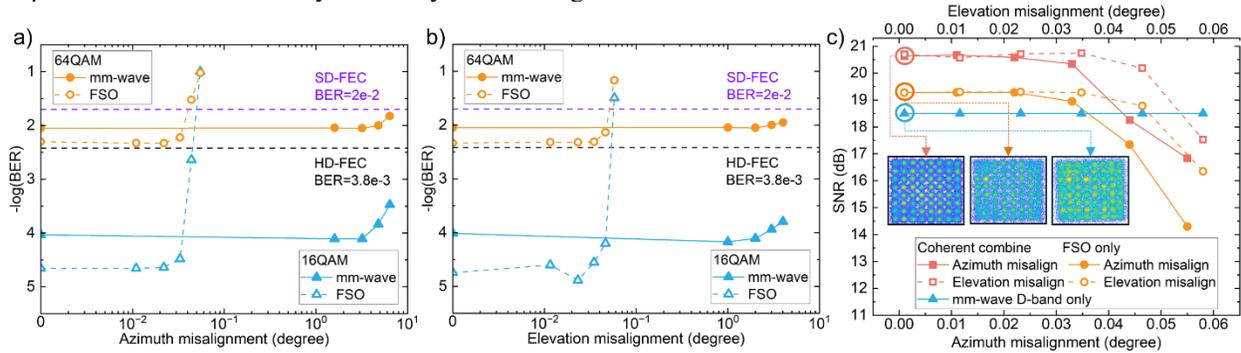

Fig. 2. BER of D-band and FSO signal with a) azimuth misalignment and b) elevation misalignment, c) SNR of coherent combined, FSO only and D-band only signal at different misalignment angles

## 4. Conclusion

We proposed and demonstrated a novel combined FSO and D-band mm-wave transmitter based on optical harmonic locking technique, enabling ultra-low phase noise D-band carrier generation and thus high quality FSO and D-band data signal transmission. With this new scheme, we demonstrated 17GBd-64QAM FSO and D-band transmission over 55cm, achieving 100Gbps per channel. We further study the misalignment tolerance of combined transceiver architecture, showing a misalignment tolerance improvement of 193 and 86 times for azimuth and elevation axis compared to standalone FSO transceiver. Finally, we have demonstrated up to 2.2dB SNR improvement by coherent combining of received D-band and FSO signal, which enables high-capacity transmission over long distance. These results indicate our combined FSO and mm-wave transceiver architecture can significantly benefit future ultra-high-capacity and high resilience wireless communications.

**Acknowledgment**
The authors would like to acknowledge funding from the EPSRC grant TRACCS (EP/W026732/1) and EU-horizon grant 6G-MUSICAL (101139176).